\documentclass[prb,twocolumn]{revtex4}

\usepackage{graphicx}
\usepackage{verbatim}
\begin{document}

\def\K{{\bf{K}}}
\def\Q{{\bf{Q}}}
\def\Gbar{\bar{G}}
\def\tk{\tilde{\bf{k}}}
\def\k{{\bf{k}}}

\title{Validity of the spin-susceptibility ``glue'' approximation 
for pairing in the two-dimensional Hubbard model}

\author{E. Khatami,$^{1,2}$ A. Macridin,$^{1}$ and M. Jarrell$^{2}$} 
\address{
$^{1}$Department of Physics, University of Cincinnati, Cincinnati, Ohio 45221, USA\\
$^{2}$Department of Physics and Astronomy, Louisiana State University, Baton Rouge, Louisiana 70803, USA}

\begin{abstract}
We examine the validity of the weak coupling spin-susceptibility ``glue''
approximation (SSGA) in a two-dimensional Hubbard model for cuprates.  For
comparison, we employ the well-established dynamical cluster approximation (DCA) 
with a quantum Monte Carlo algorithm as a cluster solver. We compare the
leading eigenvalues and corresponding eigenfunctions of the DCA and SSGA
pairing matrices. For realistic model parameters, we find that the SSGA
fails to capture the leading pairing symmetries seen in the DCA. Furthermore,
when the SSGA is improved through the addition of a term with $d$-wave
symmetry, the strength of this additional term is found to be larger
than that of the ``glue'' approximation.
\end{abstract}

\maketitle

\section{INTRODUCTION}

The pairing mechanism of cuprate superconductors has been a challenging 
problem since their discovery. To this day, among different scenarios, 
two stand out. First one is the Anderson's resonating valence bond
scenario~\cite{p_anderson_87}, in which superconductivity is pictured as 
a Mott liquid of pairs formed by the superexchange interaction~\cite{p_anderson_07}.  
This is a strong-coupling approach and can predict many features of the 
cuprate phase diagram~\cite{p_anderson_04}. Another is the body of weak-coupling 
approaches, including phenomenological models~\cite{a_millis_90}, 
fluctuation exchange~\cite{k_miyake_86,n_bickers_89} and random-phase 
approximation~\cite{d_scalapino_86}, in which the pairing interaction, i.e.,
the ``glue'', is mediated by the low-energy spin fluctuations.  

Recently, much effort has been devoted, both in experimental and in theoretical
fronts, to find more compelling evidence for the spin-fluctuation-mediated pairing. 
On the experimental side, neutron scattering data show a prominent 
peak in the structure factor at the antiferromagnetic wave vector, relevant to 
$d$-wave pairing~\cite{fong}. Using inelastic neutron scattering data to 
parametrize the effective spin-susceptibility glue interaction, 
Dahm {\em et.~al.}~\cite{dahm} find an excellent agreement between the 
numerically calculated features of the spectral function and the angle 
resolved photoemission spectroscopy data. Moreover, van Heumen 
{\em et.~al.}~\cite{heumen} find a correlation between the doping trends 
in the ``glue spectra'', derived from optical conductivity data, and the 
superconducting critical temperature.

On the theoretical side, the dynamics of this type of pairing has been 
recently investigated by numerous authors.  For instance, 
employing an extended Hubbard model, Markiewicz and Bansil~\cite{bansil}
argue that while magnetic pairing mechanism is valid in cuprates, both high 
and low energies are relevant to pairing. Similar conclusion have been drawn 
using numerical calculations of the Hubbard model which often involve the dynamical 
mean-field treatments of the smallest system relevant to $d$-wave pairing, the 
cluster of four sites~\cite{haule,t_maier_08a}. However, using similar 
techniques, others argue that only the low-frequency part of the pairing is 
important~\cite{kyung,civelli}. 

In this work, we examine the validity of the spin-susceptibility ``glue'' 
approximation (SSGA), expressed in a form similar to that of random phase approximation,
~\cite{d_scalapino_86,d_scalapino_95,n_bulut_94,t_maier_07b,t_maier_07,t_maier_08a}
\begin{equation}
\Gamma^{\text {SSGA}}(K|K')=\frac{3}{2}\bar{U}^2\chi_s(K'-K)
\label{eq:``glue''}
\end{equation}
by exploring the {\em momentum} dependence of pairing
and comparing it to the results obtained from a dynamical cluster 
approximation (DCA)~\cite{m_hettler_98,m_hettler_00,algorithm} simulation. 
Here, $\Gamma$ is the particle-particle irreducible vertex function and
$\chi_s$ is the fully dressed spin-susceptibility. $K=(\K,\omega_n)$ denotes 
both momentum and frequency and $\bar{U}$ is an effective Coulomb interaction.
Unlike most previous calculations, we employ a relatively large cluster, the 16-site 
cluster, allowing for more pairing symmetries. Note that in our 
SSGA, $\chi_s$ is also obtained from the DCA simulation. This emulates the 
use of experiment to parametrize the ``glue'' approximation~\cite{m_norman_87,dahm}.

We find that when  a finite next-nearest-neighbor hopping, $t'$, appropriate to
describe the hole-doped cuprates\cite{e_pavarini_01,a_Nazarenko_95,k_tanaka_04,e_khatami_08} 
is considered,  the SSGA form of the interaction leads to $p$-wave pairing ~\cite{footnote_Eg}, 
while for the same parameters the DCA yields robust  $d$-wave pairing. We show that 
this is due to the predominant scattering with the antiferromagnetic wave vector, 
resulting from the momentum dependence  of the spin susceptibility, and the strong 
suppression of the density of states (DOS) at the antinodal points (strong 
pseudogap)~\cite{a_macridin_06b}. To re-establish the $d$-wave pairing symmetry and the agreement 
with the DCA, the SSGA can be improved  by adding an additional term with $d$-wave 
functionality in the momentum space. However, we find that the strength of this additional 
term should be larger than that of the SSGA.

\begin{figure}[t]
\centerline {\includegraphics*[width=3.3in]{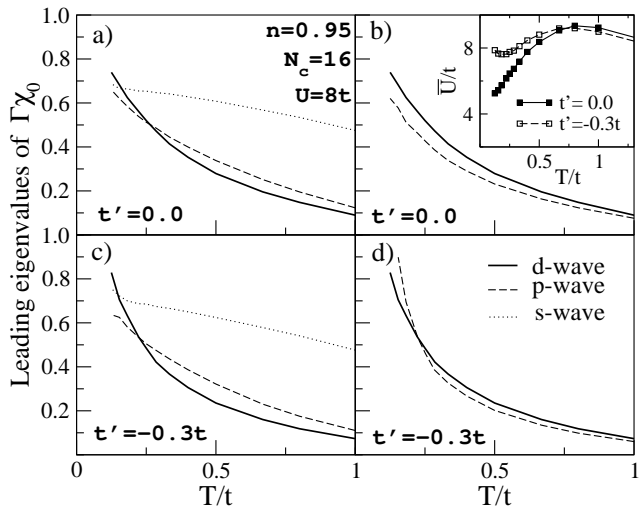}} 
\caption{Leading eigenvalues of the DCA [(a) and (c)] and SSGA [(b) and (d)] 
particle-particle pairing matrices versus temperature. Vertex function of Eq.~(\ref{eq:``glue''})
is used to form the SSGA pairing matrix. Top panels correspond 
to the Hubbard model with only nearest-neighbor hopping ($t$). In lower panels, a 
finite next-nearest-neighbor hopping is also taken into account. 
The inset of (b) shows the effective Coulomb interaction used in the SSGA vertex,
which is adjusted so that $d$-wave eigenvalue in SSGA is the same as its DCA counterpart.}
\label{fig:comparison}
\end{figure}


\section{FORMALISM}
\label{sec:formalism}

We consider a two-dimensional Hubbard Hamiltonian
\begin{eqnarray}
\label{eq:ham}
H=-\sum_{ij\sigma}t_{ij}(c^{\dagger}_{{i}\sigma}c_{{j}\sigma}+h.c.)
+U\sum_{i}n_{{i}\uparrow}n_{{i}\downarrow} 
\end{eqnarray}
where $t_{ij}$ is the hopping matrix, $c^{\dagger}_{{i}\sigma}
(c_{{i}\sigma})$ is the creation (annihilation) operator for 
electrons on site ${i}$ with spin $\sigma$ and 
$n_{i\sigma}=c^{\dagger}_{{i}\sigma}c_{{i}\sigma}$. We show results 
for $U$ equal to the bandwidth which is believed
to be a realistic value for modeling cuprates~\cite{a_mcmahan_88,e_stechel_88,m_Hybertsen_89} 
and at filling, $n=0.95$. Calculations at different hole-dopings show
that our conclusions are valid in the under-doped region ($n>0.85$),
where the antiferromagnetic correlations are stronger, while at larger dopings, 
the results are inconclusive since calculations are limited by the sign problem.

The DCA is a cluster mean-field theory that 
maps the original lattice model onto a periodic cluster of size 
$N_c=L_c^2$ embedded in a self-consistent host. Correlations up to 
a range $L_c$ are treated explicitly using a quantum Monte Carlo 
(QMC) solver, while those at longer length scales are described 
at the mean-field level. Previous DCA simulations have shown a robust 
$d$-wave superconductivity for the Hubbard model with $U$ comparable 
to the bandwidth. The DCA has also established pseudogap and 
antiferromagnetic phases for this model which are in good qualitative 
agreement with experimental phase diagram of cuprates~\cite{th_maier_05,Alex}.

To study the pairing, we calculate the eigenfunctions of the paring matrix, 
$\Gamma{\chi_0}$~\cite{n_bulut_93}, 
\begin{equation}
\frac{T}{N_c}\sum_{K'}\Gamma(K|K')\chi_0(K')\phi(K')=\lambda\phi(K)
\label{eq:Bethe}
\end{equation}
where $T$ is temperature, ${\chi_0}$ ($=-G(K)G(-K)$) is the particle-particle 
bubble in the pairing channel and $\Gamma$ can be either $\Gamma^{\text{SSGA}}$ 
[Eq.~\ref{eq:``glue''}] or calculated in the DCA. For the latter, we measure 
the two-particle Green's function in the pairing channel ($\chi$) in the QMC 
process. Then, using the Bethe-Salpeter equation, $\Gamma$ is calculated by
subtracting the inverse of $\chi$ from the inverse of the bare bubble in 
the same channel ($\Gamma=\chi_0^{-1} - \chi^{-1}$).
The eigenfunction $\phi(K)$ represents the gap function and provides information
about the symmetry and the frequency dependence of the pairing.  The singularities 
in the two-particle Green's function, ${\chi}={\chi_0}/(1-\Gamma{\chi_0})$,
which signal pairing instability, take place when the eigenvalue, $\lambda $, goes 
to unity.  Therefore, one can explore the superconducting tendencies 
by studying the temperature dependence of the leading eigenvalues.

\begin{figure}[t]
\centerline {\includegraphics*[width=3.3in]{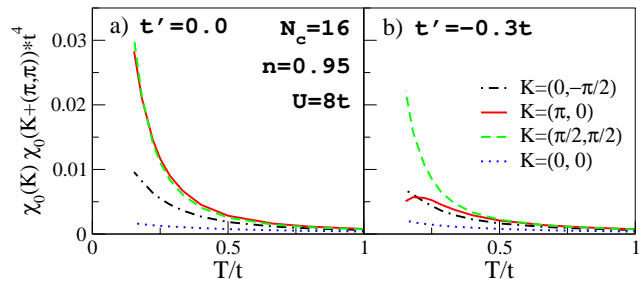}} 
\caption{(Color online) The product $\chi_0(\K)\chi_0(\K+\Q)$ with $\Q=(\pi,\pi)$ 
and $\omega_n=\omega_{n'}= \pi T$ at four different $\K$ points in the first 
Brillouin zone (1BZ) for (a) $t'=0$ and (b) $t'=-0.3t$ versus temperature. 
At low temperatures, the leading pairing symmetries in SSGA can be determined 
by the value of the product $\chi_0(\K)\chi_0(\K+\Q)$ at each $\K$ point. 
The one for $\K=(\pi,0)$ corresponds to the $d$-wave symmetry and shows a 
significant decrease with $t'$ at low temperatures.}
\label{fig:chi16B}
\end{figure}


\section{RESULTS}

Unlike DCA calculations which yield $d$-wave pairing, when a  finite next nearest  
neighbor hopping, appropriate to describe hole-doped cuprates,  is considered the 
SSGA pairing vertex alone [Eq.~\ref{eq:``glue''}] does not result in pairing with 
$d$-wave symmetry.  In Fig.~\ref{fig:comparison}, we compare the leading 
eigenvalues of the SSGA and DCA pairing matrices for zero and finite $t'$.  When 
$t'=0$, the low-temperature DCA results have a $d$-wave leading eigenvalue, followed 
by $s$-wave and $p$-wave eigenvalues. Similar results are found with the SSGA pairing 
matrix, except that $s$-wave is not one of the leading eigenvalues.  However, when 
$t'=-0.3t$, $p$-wave instead of $d$-wave becomes the dominant pairing symmetry for 
the SSGA while $d$-wave is still the leading eigenvalue in the DCA. 
Thus, the SSGA fails to capture the symmetry of the pairing obtained from the DCA 
calculation.   Here, $\bar{U}$, which is shown for the two different values 
of $t'$ in the inset of Fig~\ref{fig:comparison} (b), is adjusted so that 
$d$-wave eigenvalue in SSGA is the same as its DCA counterpart.

The momentum dependence of $\chi_s$ and the renormalization of particle-particle
bubble due to finite $t'$ are responsible for the disagreement between the 
symmetry of the leading eigenfunctions in the SSGA and the DCA.
To better understand why the dominant pairing in the  SSGA is not $d$-wave, we 
employ the following  approximation: 
$\chi_s(K'-K)\approx\chi_s(\Q,0)\delta_{\K'-\K,\Q}\delta(\omega_{n'}-\omega_n)$.
This is motivated by the fact that the spin-susceptibility 
is considerable only at the antiferromagnetic wave vector, $\Q=\K'-\K=(\pi,\pi)$
and at small Matsubara frequency. In this approximation, Eq.~(\ref{eq:Bethe}) can be 
written as
\begin{equation}
\frac{3T}{2N_c}\bar{U}^2\chi_s(\Q){\chi_0}(\K+\Q,\omega_n)\phi(\K+\Q,\omega_n)
\approx\lambda\phi(\K,\omega_n).
\label{eq:BS1}
\end{equation}
Considering that
\begin{equation}
\frac{3T}{2N_c}\bar{U}^2\chi_s(\Q){\chi_0}(\K,\omega_n)\phi(\K,\omega_n)
\approx\lambda\phi(\K+\Q,\omega_n)
\label{eq:BS2}
\end{equation}
is also true, one gets
\begin{equation}
\lambda^2\propto\chi_s(\Q)^2\chi_0(\K,\omega_n)
\chi_0(\K+\Q,\omega_n).
\label{eq:lambda}
\end{equation}
This  suggests that the leading eigenvalue of the SSGA pairing matrix corresponds 
to a momentum, $\K$, for which the quantity $\chi_0(\K,\omega_n)\chi_0(\K+\Q,\omega_n)$ 
has its largest value. Since the bubble  $\chi_0(\K,\omega_n )$ falls rapidly with 
frequency, only the lowest Matsubara frequencies are relevant for determining the 
leading eigenvalues. In the following, we will discuss  the behavior of 
$\chi_0(\K,\omega_n)\chi_0(\K+\Q,\omega_n)$ at $\omega_n=\pm \pi T$.

\begin{figure}[t]
\centerline {\includegraphics*[width=2.in]{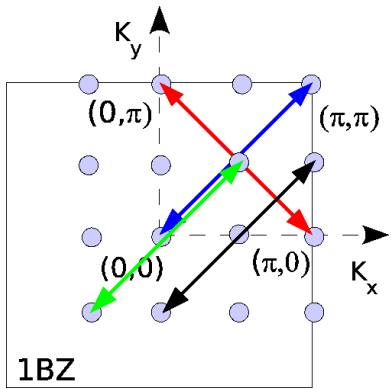}} 
\caption{(Color online) The 1BZ for $N_c=16$. Arrows show 
the four independent ($\pi,\pi$) scatterings between $\K$ and $\K'$ on this cluster. 
The one that connects ($\pi,0$) to ($0,\pi$) is associated with the $d$-wave 
symmetry.}
\label{fig:1BZ}
\end{figure}

When $t'=0$, $\chi_0(\K)\chi_0(\K+\Q)$ is the largest for $\K=(\pi,0)$ and 
$\K=(\pi/2,\pi/2)$ at low temperature, as can be seen in Fig.~\ref{fig:chi16B} (a).
The former situation favors $d$-wave pairing whereas the latter favors $p$-wave 
pairing.  The close values of $\chi_0(\K)\chi_0(\K+\Q)$ at these momenta explains 
the competition between $d$-wave and $p$-wave symmetries in SSGA. Note that for 
a small $2\times2$ cluster, the resolution in momentum space is poor (e. g. 
$\K=(\pi/2,\pi/2)$ is not represented in the Brillouin zone)
and the scattering between the nodal points which favors $p$-wave pairing
is suppressed. Thus, a larger cluster with good momentum  resolution is
important in capturing symmetries other than $d$-wave in the pairing channel.
The 16-site cluster  provides
four independent values for $\chi_0(\K)\chi_0(\K+\Q)$. This is illustrate in 
Fig.~\ref{fig:1BZ} where each arrow represents a $\Q=(\pi,\pi)$ 
scattering between $\K$ and $\K+\Q$.

A finite $t'$ strongly suppresses $\chi_0(\K)\chi_0(\K+\Q)$ at the antinodal 
points while it has a small effect at other $k$ points [see Fig.~\ref{fig:chi16B} (b)].
Therefore, according to Eq.~(\ref{eq:lambda}), the $d$-wave eigenvalue will also 
be suppressed relative to the $p$-wave eigenvalue. Consequently, the  $p$-wave
pairing will be dominant in the SSGA, which explains the results  discussed in 
Fig~\ref{fig:comparison}.  The renormalization of the bare bubble 
at the antinodal points can be understood from the changes in the DOS. At low 
temperatures, the low energy DOS at $\K=(0,\pi)$ is strongly suppressed [i.e., the 
pseudogap is enhanced] with $t'$ for the hole-doped systems while this effect is 
negligible at other $k$ points~\cite{a_macridin_06b}. This indicates that in the
over-doped region, where pseudogap is less pronounced, SSGA might be a good 
approximation.

We find that the spin-susceptibility representation of the pairing 
interaction, which is appreciable only at $\K'-\K=(\pi,\pi)$, does 
not always yield $d$-wave as the leading pairing symmetry. This 
seems to be especially true when realistic parameters for cuprates such 
as next-nearest-neighbor hopping are considered. We show that despite 
the inherent correspondence between the  $\Q=(\pi,\pi)$ wave vector, at which 
the SSGA interaction is large, and the $d$-wave symmetry, other pairing 
symmetries which also associate with the $(\pi,\pi)$ wave vector can 
be dominant. On the other hand, the DCA exhibits a robust $d$-wave 
pairing for the same physical parameters of the Hubbard model.

\begin{figure}[t]
\centerline {\includegraphics*[width=2.8in]{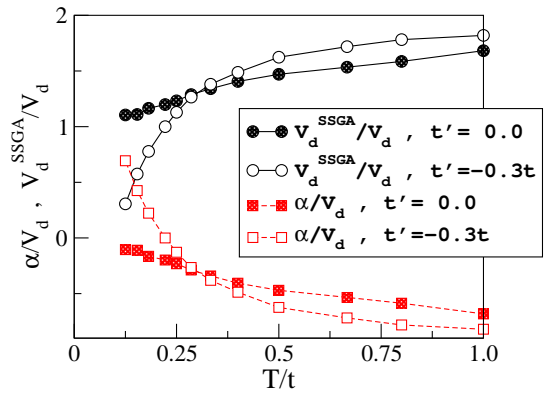}} 
\caption{(Color online) Fractions of the $d$-wave pairing interaction for the SSGA 
term (first term) and the $\alpha$ term (second term) of Eq.~(\ref{eq:glue2}) for zero 
and finite $t'$ versus temperature. $V_d=V_d^{\text{SSGA}}+\alpha$ is the total
$d$-wave projected interaction of Eq.~(\ref{eq:glue2}).
For $t'=0$, the contribution of the $\alpha$ term to the $d$-wave interaction
is insignificant while for $t'=-0.3t$, it becomes more important than the 
SSGA term at low temperatures.}
\label{fig:alphaVd}
\end{figure}

To investigate the possibility of a missing term, we propose to add an extra
term to $\Gamma^{\text{SSGA}}$ which enhances the scattering between the antinodal points.
Such approximation for the interaction vertex can be written as:
\begin{equation}
\Gamma'(K|K')=\frac{3}{2}\bar{U}^2\chi_s(K'-K)-\alpha \phi_d(K')\phi_d(K).
\label{eq:glue2}
\end{equation}
Here, $\phi_d$ is the $d$-wave eigenfunction of the DCA pairing matrix, and $\bar{U}$ and  
$\alpha$ are temperature-dependent fitting parameters which are adjusted to reproduce 
both the $d$-wave and the $p$-wave leading eigenvalues of the DCA pairing matrix.  It 
is worth mentioning that Maier {\em et. al.}~\cite{t_maier_08a} proposed a similar 
term, $-\bar{J}g(\K)g(\K')$, to be added to the SSGA form in which 
$g(\K) \propto (\cos K_x-\cos K_y)$ is the $d$-wave form factor and $\bar{J}$ is an 
effective exchange interaction.  However, their motivation for the necessity of this extra 
term is quite different from ours. Since they did not consider a finite $t'$, the SSGA gave 
a $d$-wave pairing. So, this term was suggested only to restore the large frequency 
behavior of the pairing gap found in the DCA. Our focus, on the other hand, is to examine 
the relevance of an additional term based on a momentum space argument. Note that because our 
calculations are done using QMC on the imaginary frequency axis, generally comments about the 
frequency dependence of the pairing interaction cannot be made. However, if relevant,
the additional term may be responsible for the instantaneous part of the interaction,
as described in Ref.~\onlinecite{t_maier_08a}. We also find that using $\phi_d(K)$ 
instead of $g(\K)$ in our approximation provides a better fit to the DCA results while imposing 
the $d$-wave symmetry.

For a finite $t'$, the second term in Eq.~(\ref{eq:glue2}) ($\alpha$ term) has a 
prominent role in capturing the correct leading symmetries in this new form of the 
interaction.  In Fig.~\ref{fig:alphaVd}, we show the fractional values of $\alpha$ 
and the $d$-wave component of the SSGA term~\cite{t_maier_06b,t_maier_07b}, 
\begin{equation}
\label{eq:1}
V_d^{\text{SSGA}}=-\sum_{K,K'}\phi_d(K')\frac{3}{2}\bar{U}^2\chi_s(K'-K)\phi_d(K),
\end{equation}
for the two values of $t'$.
We define $V_d=V_d^{SSGA}+\alpha$ as the total $d$-wave projected interaction.  When 
$t'=0$, the contribution of the $\alpha$ term to the $d$-wave interaction is 
insignificant. However, when $t'=-0.3t$, this term has a dominant role in the $d$-wave 
pairing at low temperatures. So, although the additional term enhances $d$-wave
pairing, its contribution overshadows the contribution of the main part of the interaction 
i.e. the term proportional to the spin-susceptibility. This suggests that the SSGA
may not be improved by simply adding terms that enhance $d$-wave scattering.

\section{CONCLUSIONS}

We study the validity of the SSGA representation of 
the pairing interaction in the Hubbard model. By comparing the leading pairing 
symmetries of this interaction with the pairing symmetries produced by the DCA, 
we find that this approximation alone does not capture the correct pairing 
symmetry in the under-doped and optimally-doped regions, particularly when a 
finite $t'$ is considered. We do not dismiss the possibility that SSGA can be valid 
in the over-doped region, where the pseudogap phenomenon is less relevant.
We show that this form of the interaction, which is large only at the 
antiferromagnetic wave vector, requires an additional term with $d$-wave symmetry 
to yield $d$-wave pairing at low temperatures. However, in case of 
finite $t'$, the additional term dominates the interaction as the temperature 
is lowered. 

\section{ACKNOWLEDGMENTS}

This research was supported by NSF under Grant No. DMR-0706379, and DOE CMSN 
under Grant No. DE-FG02-04ER46129 and enabled by allocation of advanced computing 
resources, supported by the National Science Foundation. The computations 
were performed on Lonestar at the Texas Advanced Computing Center
(TACC) under Account No. TG-DMR070031N.


\begin{thebibliography}{99}
\bibitem{footnote_Eg} In this paper, by $p$-wave symmetry we denote the doubly degenerate
$E_g$ irreducible representation of the $D_{4h}$ point group associated with the square lattice.
\end{thebibliography}
\end{document}